\documentclass[journal]{IEEEtran}

\ifCLASSINFOpdf
\else
   \usepackage[dvips]{graphicx}
\fi
\usepackage{url}

\usepackage{graphicx}
\usepackage{amsfonts}
\usepackage{mathrsfs}
\usepackage{dutchcal}
\usepackage{amsmath}
\usepackage{booktabs}
\usepackage{threeparttable}
\usepackage{multirow}
\usepackage[numbers,sort&compress]{natbib}
\usepackage{subfigure}
\usepackage{balance}
\usepackage{xcolor}
\begin{document}
\abovedisplayshortskip=0.6pt
\belowdisplayshortskip=0.6pt
\abovedisplayskip=0.6pt
\belowdisplayskip=0.6pt

\title{HDRfeat: A Feature-Rich Network\\ for High Dynamic Range Image Reconstruction}

\author{Lingkai Zhu, Fei Zhou, Bozhi Liu, Orcun Goksel
\thanks{L.\,Zhu and O.\,Goksel are with the Department of Information Technology, Uppsala University, Sweden. 
F.\,Zhou and B.\,Liu are with the College of Electronic and Information Engineering, Shenzhen University, China.
This work was performed during B.\,Liu's visit at Uppsala University, Sweden.}
}

\markboth{Zhu \MakeLowercase{\textit{et al.}}: HDR\MakeLowercase{feat}: A Feature-Rich Network for HDR Image Reconstruction}
{Zhu \MakeLowercase{\textit{et al.}}: HDR\MakeLowercase{feat}: A Feature-Rich Network for HDR Image Reconstruction}
\maketitle

\begin{abstract}

A major challenge for high dynamic range (HDR) image reconstruction from multi-exposed low dynamic range (LDR) images, especially with dynamic scenes, is the extraction and merging of relevant contextual features in order to suppress any ghosting and blurring artifacts from moving objects. 
To tackle this, in this work we propose a novel network for HDR reconstruction with deep and rich feature extraction layers, 
including residual attention blocks with sequential channel and spatial attention.
For the compression of the rich-features to the HDR domain, a residual feature distillation block (RFDB) based architecture is adopted. 
In contrast to earlier deep-learning methods for HDR, the above contributions shift focus from merging/compression to feature extraction, the added value of which we demonstrate with ablation experiments.
We present qualitative and quantitative comparisons on a public benchmark dataset, showing that our proposed method outperforms the state-of-the-art.  

\end{abstract}


\IEEEpeerreviewmaketitle

\section{Introduction}

Dynamic range is the luminance ratio between the brightest and darkest areas in a scene. 
Natural scenes have a much higher luminance range than digital cameras can represent, resulting in captured low dynamic range (LDR) images to often have over- or under-exposed regions~\cite{reinhard2010high}. 
High dynamic range (HDR) images can represent a wide range of illuminations. 

The typical way to generate an HDR image is to merge a series of LDR images with different exposures of the same scene captured by a positionally-fixed camera. 
Given the acquired LDR images, some methods focus on recovering the camera response function~\cite{debevec2008recovering,mann1994being,granados2010optimal,reinhard2010high,yan2017high,salih2012tone}, which is the relation between the irradiance map and the HDR-image pixel values.
Others adopt a multiple exposure fusion approach~\cite{jinno2012multiple, pradeep2012robust,xu2022multi,zhang2021benchmarking}, which combines the multi-exposed LDR images to estimate the irradiance value for each pixel. 
However, if the input LDR images contain large foreground motions, the merged HDR image will suffer from ghosting and blurring artifacts due to the misalignment among the LDR images. 

\begin{figure}
\centering{\includegraphics[width=\linewidth]{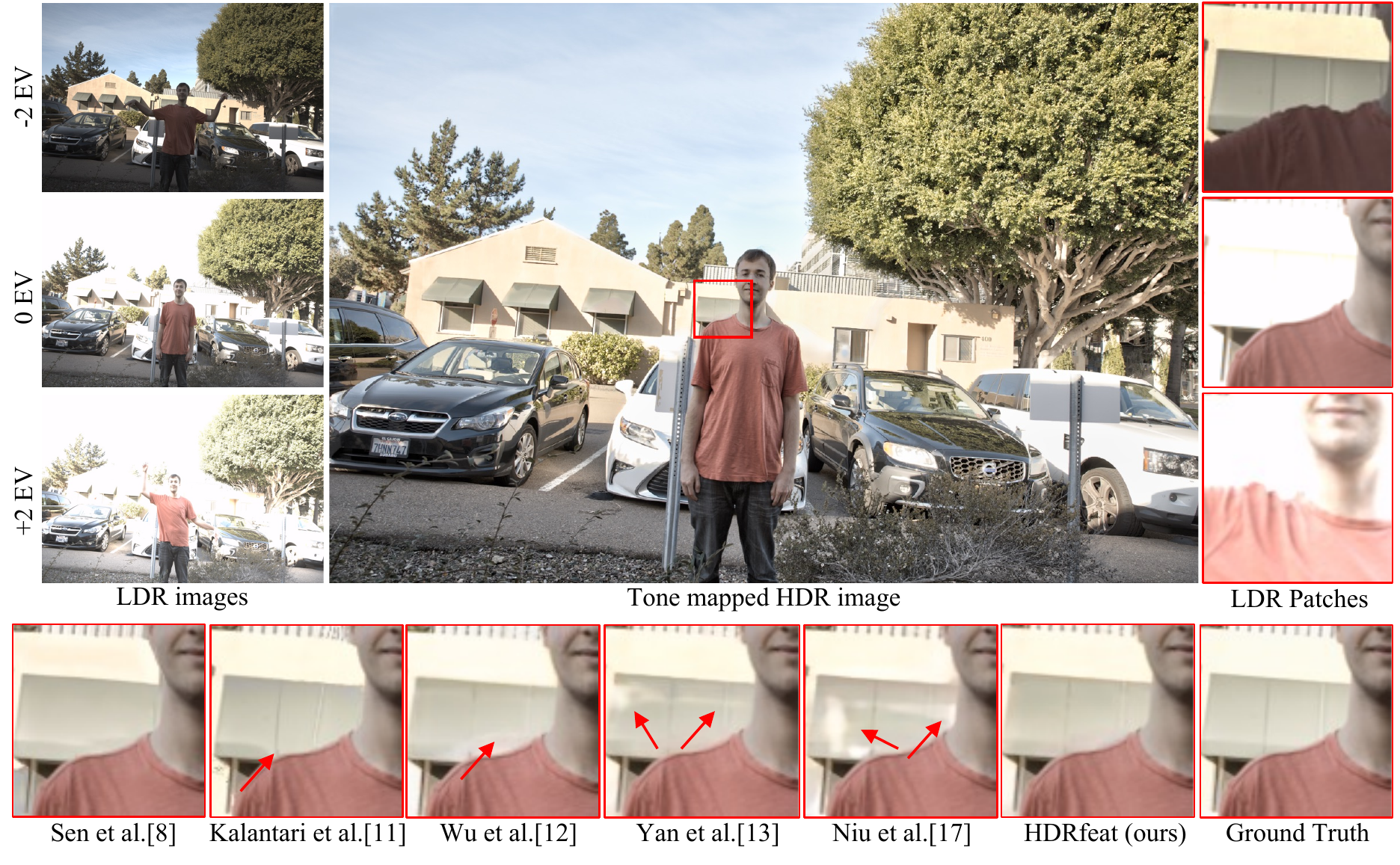}}
\caption{Visual results on the test dataset~\cite{kalantari2017deep}. The top row shows three input LDR images, the tone-mapped HDR image, and sample LDR patches at the same image location. The bottom row compares the same patch from HDR images reconstructed by state-of-the-art methods and our proposed method HDRfeat. The arrows highlight some artifacts, discussed later in our Results.}
\label{fig:visual_comparison_first}
\end{figure} 

Several deep learning approaches have been proposed to address the aforementioned problems~\cite{kalantari2017deep,wu2018deep,yan2019attention,yan2019multi,yan2022dual,wang2021exposure, niu2021HDRGAN}. 
For example, Kalantari et al.~\cite{kalantari2017deep} used optical flow to align input images and then merged the aligned images with a four-layer fully convolutional neural network (CNN) to generate a restored HDR image. 
However, such a simple neural network structure cannot handle the artifacts and distortion caused by errors from an unreliable optical flow. 
Wu et al.~\cite{wu2018deep} proposed the first end-to-end deep neural network without optical flow alignment for HDR reconstruction, which includes three encoder networks, a merger network, and a decoder network. 
However, due to the lack of an explicit alignment mechanism among LDR images, this approach suffers from occlusion and ghosting. 
In their seminal work, Yan et al.~\cite{yan2019attention} were the first to use attention networks (AHDRNet) aiming to highlight or align large motions in the features prior to the merger network, which can then suppress any ghosting artifacts.
Besides the attention network dramatically increasing the computational cost, there still existed blurring or artifacts in large motion scenarios as shown in Fig. \ref{fig:visual_comparison_first}.
Nevertheless, this work inspired several following works, which adapted their proposed architecture; for example, see several solutions in the recent NTIRE 2021 and 2022 challenges on HDR~\cite{eduardo2021ntire, perez2022ntire}.
In \cite{yan2022dual}, Yan et al.\ proposed DAHDRNet with recursive channel and spatial attention for more effective de-ghosting. 
Most deep-learning solutions to HDR (e.g.\ AHDRNet and its derivatives) involve two consecutive stages overall: extraction of features from LDR images, and based on these merging/compression into an HDR image.
The latter stage in essence is a function of LDR images, e.g.\ how to scale and merge them locally.
It is the features from the former stage that parametrizes such merging function, e.g.\ that should highlight/suppress misaligned regions, occlusions, etc.
In other words, one can expect the former task to be far more complex than the latter task.
Based on this motivation, we propose a network architecture (HDRfeat) which is rich in feature extraction with attention and channel-wise hierarchical layers, while lighter on the latter reconstruction side, compared to most state-of-the-art solutions.
Accordingly, our main contributions herein include: 
$\bullet$~a novel channel-wise hierarchical feature extraction network to efficiently extract rich contextual information from multi-exposure images, with a channel bottleneck structure to compress these LDR feature representations purposed for subsequent HDR reconstruction;
$\bullet$~a residual attention block with sequential channel and spatial attention to allow the features to focus where needed, e.g.\ large motions;
%
%
%
$\bullet$~the state-of-the-art on HDR reconstruction, demonstrated via extensive evaluations on a public benchmark dataset.

\section{Methods}

\begin{figure*}
\centering{\includegraphics[width=0.95\linewidth]{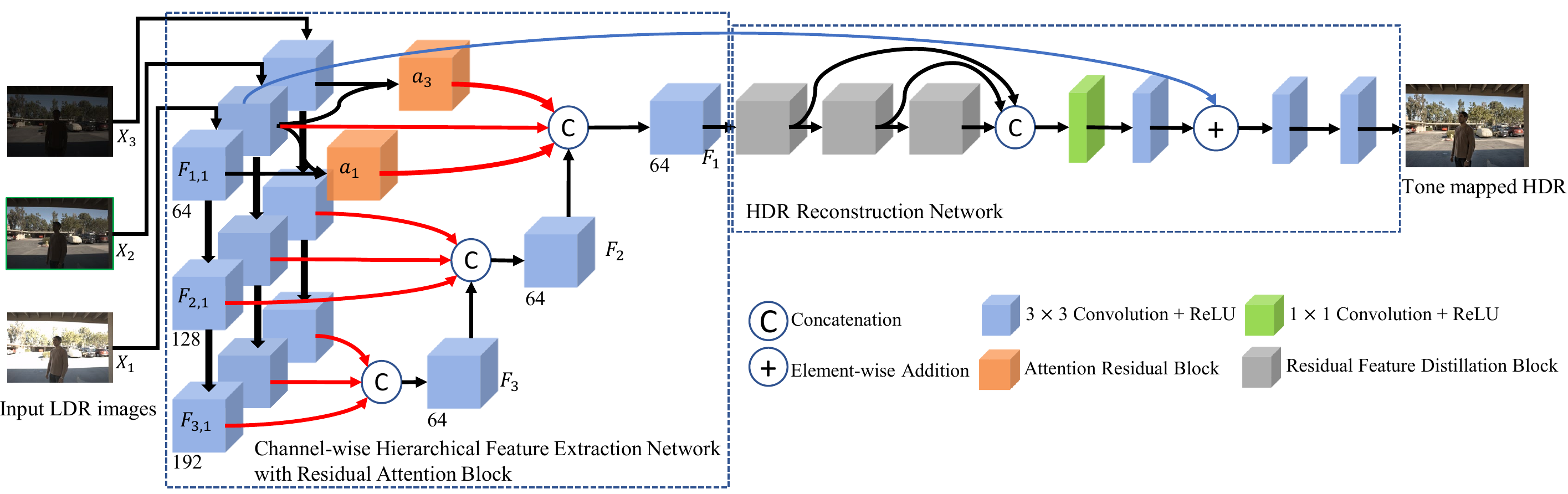}}
\caption{Proposed network architecture, involving sub-networks for hierarchical rich feature extraction with sequential attention blocks and HDR reconstruction via residual feature distillation. The former sub-network applies channel-wise hierarchical processing (in contrast to spatial hierarchy/level-of-detail, e.g., as in~\cite{niu2021HDRGAN}) together with two sequential (channel-spatial) attention blocks with residual connections and shared weights to learn merging information from non-reference exposures. The reconstruction sub-network inspired by \cite{yan2019attention, yan2022dual} uses the merged features and the first-level feature of the medium-exposure image to generate an HDR image via three consecutive residual feature distillation blocks (RFDBs) inspired by~\cite{liu2020residual} in addition to a long skip-connection that helps preserve and incorporate early features closer to the original input image.}
\label{fig:ournetwork}
\end{figure*}

Let $\{I_1, I_2, I_3\}$ be a set of 3 LDR images denoting short, medium, and long exposures, respectively.
Following~\cite{kalantari2017deep}, we preprocess $\mathcal{I}$ by gamma correction ($\gamma$ = 2.2 herein) to generate a set of mapped images $\{H_1, H_2, H_3\}$, which
concatenate with the LDR images, i.e.\ $X_i = \text{concat}(I_i, H_i)$ as input to our network.
Accordingly, an HDR image $H$ is restored by the proposed HDRfeat network $h(\cdot)$ as 
\begin{gather}
    H = h(X_1, X_2, X_3; \theta)
\end{gather}
where $\theta$ are the network parameters.
We train this network for visual acuity using a loss on tone-mapped images~\cite{kalantari2017deep} as
\begin{gather}
    \mathcal{L} = \Vert \mathcal{T}(H) - \mathcal{T}(\hat{H}) \Vert_1
\end{gather}
where $H$ and $\hat{H}$ are the predicted and groundtruth HDR images, and $\mathcal{T}(H) = \frac{\log(1+\mu H)}{\log(1+\mu)}$ is the $\mu$-law tone mapping with $H$ scaled within $[0, 1]$ and $\mu$ = 5000 in this work.

\subsection{Channel-wise Hierarchical Feature Extraction Network}

As the input LDR images contain rich contextual information, e.g.\ about foreground motions and the scene background, simple layers such as in~\cite{yan2019attention} may not be sufficient to effectively extract such information. 
In the GAN-based solution of~\cite{niu2021HDRGAN}, a spatially multi-scale deep encoder was proposed to extract visual feature from LDR images.
Nevertheless, such spatial down- and upsampling may lose the fine alignment information required in the subsequent reconstruction stage.
Furthermore, this operation is limited in enriching features in the channel dimension, which is potentially more relevant.
Therefore, inspired by the spatially hierarchical feature extraction in the generator of~\cite{niu2021HDRGAN}, we herein introduce a channel-wise hierarchical rich feature extractor (Fig.\,2) and incorporate this in a CNN solution, which is more stable to train compared to GANS.
For channel-wise hierarchy, we extract features at the same scale with expanding channel sizes, which are then merged at bottlenecks for channel dimensions to create summary features for the reconstruction. 

Architecturally, each depth of channel-wise feature extractor is composed of three shared convolution operators $E$ (3$\times$3 Conv with zero-padding of 1 pixel), extracting different levels of information denoted here as $F_{i, j} (i, j \in \{1, 2, 3\})$ with $i$ the hierarchical depth and $j$ the LDR index, i.e.:
\begin{align}
F_{1, j} & = E_{1, j}(X_j) \\ 
F_{2, j} & = E_{2, j}(F_{1, j}) \\
F_{3, j} & = E_{3, j}(F_{2, j})
\end{align}
with respective dimensions of 64$\times$128$\times$128, 128$\times$128$\times$128, and 196$\times$128$\times$128. 

In the spirit of several works that demonstrated the value of attention in HDR, we also adopt this strategy.
To that end, we introduce convolutional block attention modules (CBAM) with residual attention for the first time in HDR reconstruction, as described in detail in the following subsection.
Ultimately, our attention residual blocks $A_1$ and $A_3$ help introduce attention from the medium-exposure features $F_{1,2}$ as reference, into the short- and long-exposure features $F_{1,1}$ and $F_{1, 3}$, generating the feature maps
\begin{gather}
    a_i = A_i(\,F_{1, i}\,,\,F_{1, 2}\,), \qquad i \in \{1, 3\}\,. 
\end{gather}

The rich contextual information (denoted by the red lines in Fig.\,2) are next compressed via sequential concatenation and (3$\times$3) convolutions, to obtain summary feature representations $F_j$ at each hierarchical depth $j$ as follows:
\begin{align}
 F_3 &= \text{Conv}\big(\text{Concat}(F_{3, 1}, F_{3, 2}, F_{3, 3})\big) \\
 F_2 &= \text{Conv}\big(\text{Concat}(F_{2, 1}, F_{2, 2}, F_{2, 3}, F_3)\big) \\
 F_1 &= \text{Conv}\big(\text{Concat}(a_1, F_{1, 2}, a_3, F_2)\big) 
\end{align}
all with dimensions 64$\times$128$\times$128.

\subsection{Residual Attention Block}
Attention mechanisms have been shown to be beneficial in several deep learning tasks, such as object detection \cite{Woo2018cbam} and super-resolution \cite{zhang2018residual}, as well as HDR reconstruction~\cite{yan2019attention, yan2022dual}. 
Although spatial and channel attention can both be instrumental and can help reduce ghosting artifacts~\cite{yan2019attention, yan2022dual}, their introduction using traditional parallel layer branching, as in~\cite{yan2019attention, yan2022dual}, requires high computational resources~\cite{wang2021deep}.
Furthermore,the multiplication of such multiple attention branches makes the gradients in general unstable to effectively back-propagate during training. 
To address these, we adopt a sequential attention mechanism, inspired by Convolutional Block Attention Module (CBAM)~\cite{Woo2018cbam}, where first the global information is put in focus via channel attention and then the local information is explored via spatial attention. 
Thus, such attention mechanism operates in a global-to-local manner.
In contrast to~\cite{Woo2018cbam}, which involves a single image input (for object detection purposes), we wish to inject attention with the help of one image (reference) into another (target).
Accordingly, we first concatenate both image features as our block input, and after the addition of attention, we bring back the target image as a residual input, such that the attention will only need to robustly learn the difference from the target (see Fig.\,3).
\begin{figure}
\centering{\includegraphics[width=0.75\linewidth]{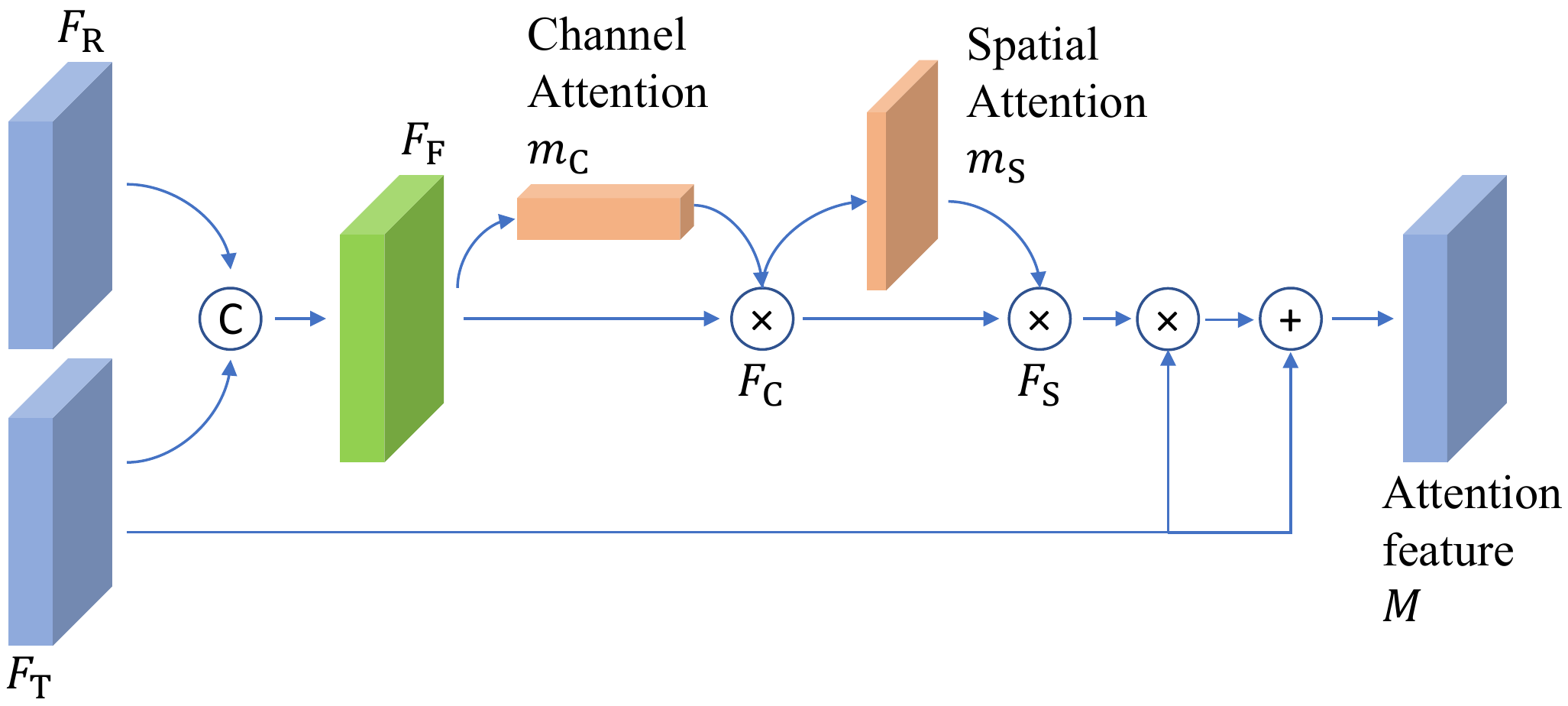}}
\caption{Our attention module brings in attention from the medium-exposed (reference) image features $F_\mathrm{R}$ into the short- or long-exposed image (target) feature $F_\mathrm{T}$, by first fusing them (concatenation + 1$\times$1 convolution) into $F_\mathrm{F}$, then finding channel and spatial attention sequentially, then applying this on the target features, and finally adding target features as a residual input. }
\label{fig:light-weight attention}
\end{figure}
For a formal definition, let the medium-exposed image (reference) features $F_{1,2}$ be denoted as $F_\mathrm{R}$, and the short- or long-exposed image (target) features, $F_{1,1}$ or $F_{1,3}$, as $F_\mathrm{T}$.
We then find the final attention applied features $M\in[0, 1]$, with the same dimension as $F_R$ and $F_{T}$, as follows:
\begin{align}
    M &= a(F_\mathrm{R}, F_\mathrm{T})\\
    M &= F_\mathrm{T} + F_\mathrm{T} \otimes F_\mathrm{S}\\
    F_\mathrm{S} &= F_\mathrm{C} \otimes m_\mathrm{S}(F_\mathrm{C})\\
    F_\mathrm{C} &= F_\mathrm{S} \otimes m_\mathrm{C}(F_\mathrm{I})\\
    F_\mathrm{F} &= \text{Conv}(\text{Concat}(F_\mathrm{R}, F_\mathrm{T}))
\end{align}
where the operator $\otimes$ denotes the element-wise multiplication; $F_\mathrm{F}$, $F_\mathrm{C}$, and $F_\mathrm{S}$ denote features for fused features, spatial attention, and channel attention;  Conv is a 1$\times$1 convolution; and $m_\mathrm{C}$ and $m_\mathrm{S}$ are core attention modules with small light-weight CNNs for low computation cost.
Channel attention $m_\mathrm{C}$ of dimension 64 first calculates the average-pooled feature and max-pooled feature separately, then forwards them to a multilayer perceptron (MLP) with one hidden layer. Next, the two output features of MLP are merged via element-wise summation and gated by a sigmoid function.
Spatial attention $m_\mathrm{S}$ of dimension 128$\times$128 first calculates the average-pooled feature and max-pooled feature separately, which then will be concatenated, forwarded to a 7$\times$7 Conv layer, and gated by the sigmoid function.

Intuitively, max-pooling extracts the most salient features that are the features with large foreground motions while the average-pooling extract the global features. By combining both features, our attention module is able to concentrate on features with large object movement in the global scenery. 
\subsection{HDR Reconstruction Network for HDR Image Restoration}
We adapt the HDR reconstruction architecture from \cite{yan2019attention, yan2022dual}.
Since, given suitable features as input, this latter reconstruction processing stage should be intuitively far less complex compared to the former feature extraction, we propose the replacement of the original cumbersome dilated residual dense blocks (DRDBs) with a more light-weight structure.
To that end, we adopt residual feature distillation blocks (RFDBs) proposed in~\cite{liu2020residual} for image super-resolution.
We extend RFDB by incorporating dilated convolutions in order to capture larger context for our purposes of HDR reconstruction.
Accordingly, our RFDB adaptation consists of 2-dilated convolution layers, shallow residual blocks and enhanced spatial attention (ESA) blocks.
In our preliminary experiments, we observed this reconstruction architecture to have performance comparable to DRDB, but with much smaller number of parameters, which we then use to invest in our rich-feature extraction network.
Since training such networks with reasonable batch sizes takes a large part of typical GPU memory, designing and investing parameters (degrees-of-freedom) within an overall network architecture in line with the particular problem structure is therefore fundamental in preventing overfitting and making best use of available resources.
Accordingly, our proposed shift of network complexity from reconstruction to feature extraction is a major contribution of this work, and it is a main component that makes HDRfeat competitive and superior to state-of-the-art.
\section{Experiments and Results}
\label{section: experiments}

\subsection{Experimental Setting and Details}

We train our proposed model on Kalantari's HDR dataset \cite{kalantari2017deep}, which contains 74 images for training and 15 images for evaluation, all with corresponding LDR and HDR images (of resolution 1500$\times$1000) as groundtruth. 
Our proposed method was implemented in Pytorch and trained on a RTX\,3090 GPU with 16\,000 epochs. 
For training, instead of feeding entire images into the network, which would require unattainable amount of memory, we randomly crop images into 256 $\times$ 256 patches. 
We apply data augmentation (flips and rotations) during training. 
We use the Adam optimizer \cite{kingma2015AdamAM}, with a learning rate initialized to $10^{-4}$ for all layers and decreased to $10^{-5}$ after 12\,000 epochs. 
All convolution layer weights are initialized with the Kaiming method~\cite{he2015delving}.
For evaluation, we compute the PSNR and SSIM scores between the restored HDR image and the groundtruth HDR image in the linear domain (PSNR-L, SSIM-L) as well as after tone mapping using $\mu$-law (PSNR-T, SSIM-T). 
We also compute the HDR-VDP-2 score \cite{mantiuk2011hdr}. 

\subsection{Comparison with state-of-the-art}

We compare our proposed method with state-of-the-art approaches, including a patch-based technique~\cite{pradeep2012robust}, an optical-flow based method with a CNN merger~\cite{kalantari2017deep}, as well as four other well-known deep-learning based methods~\cite{wu2018deep, yan2022dual, yan2019attention, niu2021HDRGAN}. 
All methods are reproduced using the codes provided by the authors, except for the results of \cite{yan2022dual} reported directly from their paper.
The comparison is presented in Table~\ref{table: comparison1}.
\begin{table}
\renewcommand{\arraystretch}{0.9}
\setlength\tabcolsep{3pt}
\centering
\caption{Quantitative comparison of our method on the test set~\cite{kalantari2017deep}. }
\label{table: comparison1}
\begin{tabular}{c c  c  c  c  c } 
    \toprule
& PSNR-T & PSNR-L & SSIM-T & SSIM-L &HDR-VDP-2  \\
    \midrule
    Sen et al.\cite{pradeep2012robust} &40.95  & 38.27  & 0.9858 & 0.9762 & 61.72 \\
    Kalantari et al.\cite{kalantari2017deep} & 42.67 & 41.21& 0.9889 & 0.9829 & 65.01\\
    Wu et al.\cite{wu2018deep} & 42.70 & 41.13 & 0.9910 & 0.9889 & 66.20 \\
    Yan et al.\cite{yan2019attention} & 43.62 & 41.03 & 0.9919 & 0.9887 & 65.79\\
    Yan et al.\cite{yan2022dual} & 43.84 & 41.31& - & - & -   \\
    Niu et al.\cite{niu2021HDRGAN} & 43.92 & 41.57  & 0.9925 & 0.9898 & 65.81 \\
    HDRfeat (ours)  & \textbf{44.11} & \textbf{41.79} & \textbf{0.9931} & \textbf{0.9912} & \textbf{66.74} \\
    \bottomrule
\end{tabular}
\end{table}

\begin{figure}
\centering{\includegraphics[width=\linewidth]{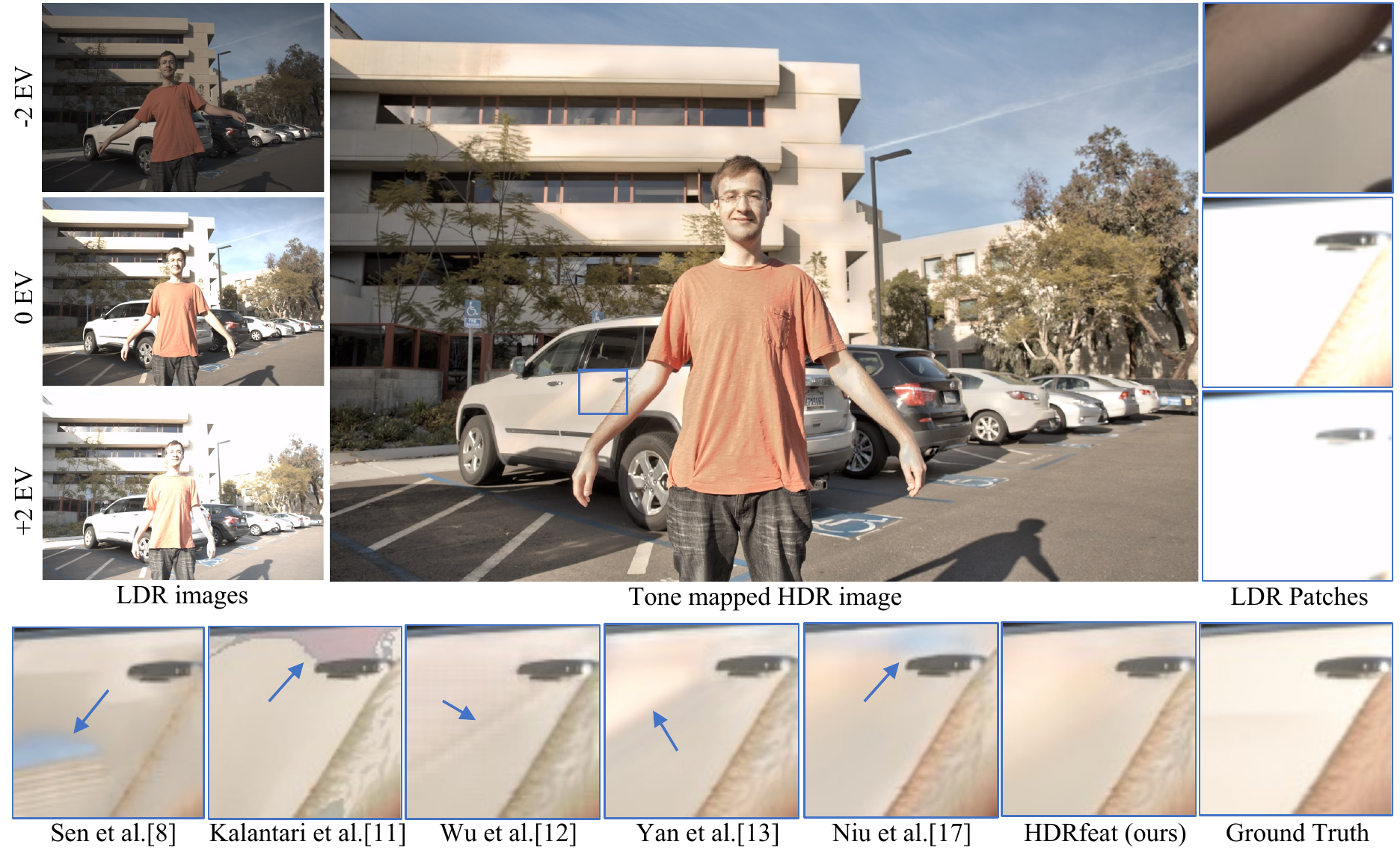}}
\caption{Visual comparison on~\cite{kalantari2017deep}. 
The arrows highlight some artifacts in the generated HDR images.}
\label{fig:visual_comparison}
\end{figure}

The result shows that our method outperforms the state-of-the-art techniques on all metrics.
In general, our method is seen to produce high quality output with less artifacts in saturated and motion-involved (occluded) image regions.
Fig.\,1 shows a visual comparison of the tone mapped HDR images, where a saturated image region is shown as zoomed-in: 
As shown with the arrows in the images, \cite{pradeep2012robust} and \cite{kalantari2017deep} show superfluous contrast, and \cite{wu2018deep}, \cite{yan2019attention}, and \cite{niu2021HDRGAN} show quantization artifacts, while our results are superior visually.
Note that no visual comparison could be included from~\cite{yan2022dual} since the code was not available at the time of this work.
Fig.\,\ref{fig:visual_comparison} shows an example, where the zoomed in location shows the car door that is occluded by the moving arm in one exposure:
As seen, our proposed method has minimal to no artifacts, whereas \cite{pradeep2012robust} and \cite{kalantari2017deep} show irrelevant content; \cite{wu2018deep} and \cite{yan2019attention} show artifactual boundaries from occluding arm; and \cite{niu2021HDRGAN} shows some leakage.

\begin{figure}
    \centering
    \includegraphics[width=0.8\linewidth]{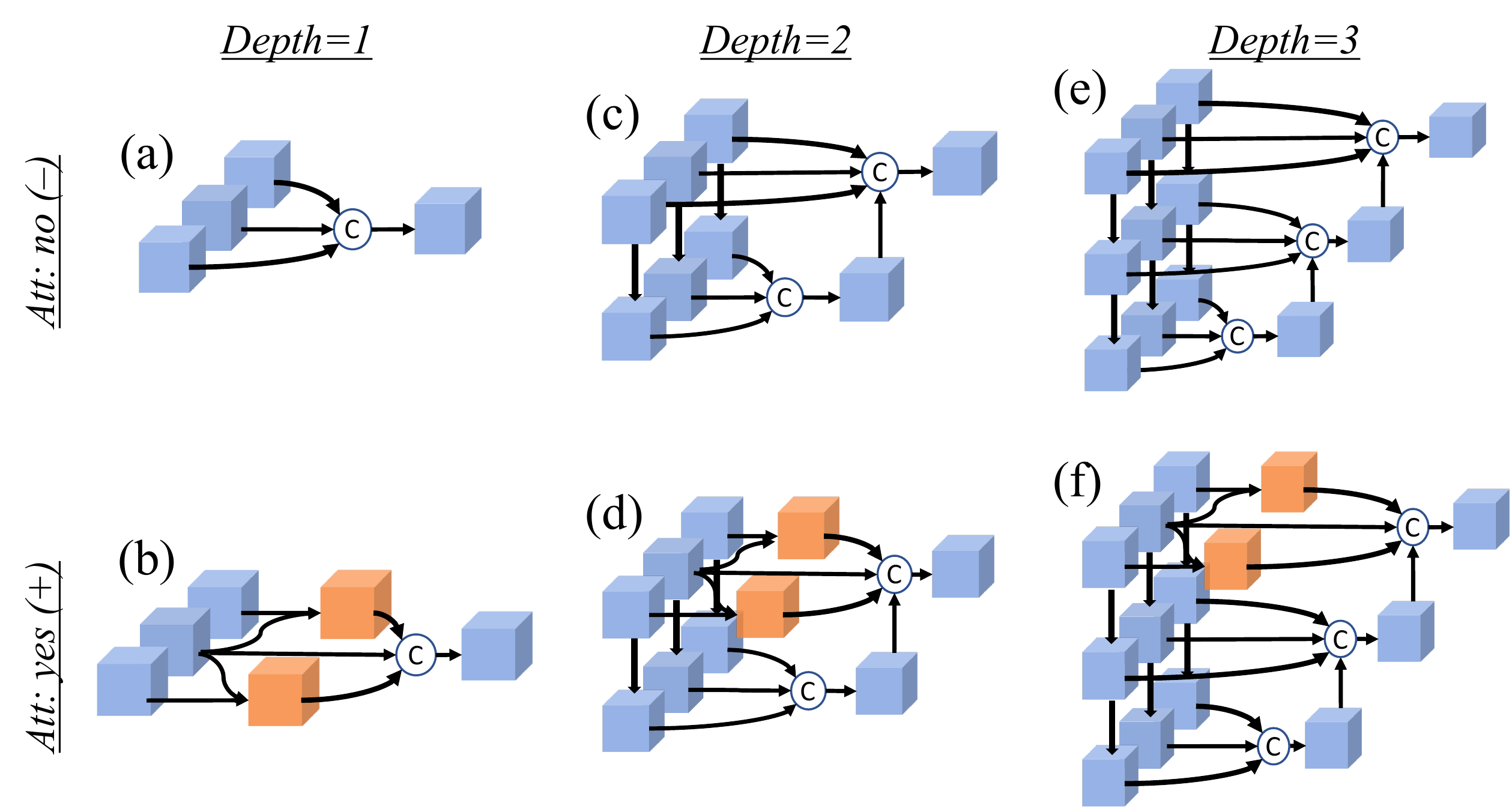}
    \caption{Architectures in our ablation study of feature extraction network, with varying depths and with/without attention modules.}
    \label{fig:variants}
\end{figure}

\subsection{Ablation study}

We investigate the components of our proposed hierarchical rich feature extraction, by ablation the attention architecture as well as the hierarchical depth (richness) of the feature extraction.
We accordingly compare our proposed HDRfeat with its ablated variants having different feature extraction depths and enabled/disabled attention modules, as shown in Fig.\,5.
In addition, we compared our sequential arrangement of spatial and channel attention, to its parallel counter-part~\cite{Woo2018cbam}, within the same HDRfeat structure  as in Fig.\,5(f).
Results are tabulated in Table \ref{table:comparison_ablation}.
\begin{table}
\renewcommand{\arraystretch}{0.9}
\setlength\tabcolsep{3pt}
\centering
\caption{Ablation study, where Depth indicates the hierarchical feature extraction depth as an indicator for the richness of features, Att indicates whether attention is applied (+) or not~(--), where +$^*$ indicates the use of parallel attention.}
\label{table:comparison_ablation}
\begin{tabular}{c | c || c c c c c} 
    \toprule
      Depth & Att & PSNR-T & PSNR-L & SSIM-T & SSIM-L & HDR-VDP-2  \\
    \midrule
    \multirow{2}*{1} & -- & 43.58 & 41.26 & 0.9924 & 0.9903 & 65.54\\
    ~ & + & 43.44 & 40.94 & 0.9924 & 0.9902 & 65.08\\
     \midrule
    \multirow{2}*{2} & -- & 43.38 & 41.24 & 0.9926  & 0.9905  & 65.54\\
    ~ & + & 43.78 & 41.61 & 0.9926 & 0.9902 & 66.38\\
    \midrule
     \multirow{2}*{3} & -- & 43.52  & 41.62 & 0.9928 &  0.9906 & \textbf{66.79}\\
    ~ & + & \textbf{44.11} & \textbf{41.79} & \textbf{0.9931} & \textbf{0.9912} & 66.74\\
    \midrule
     3 & \,\ +$^*$ & 43.86 & 41.33 & 0.9928 & 0.9903 & 65.72 \\
    \bottomrule
\end{tabular}
\end{table}

As seen, both the attention modules and the hierarchical rich-feature extraction contribute to the results, with their combination as proposed achieving the best results for most (4 out of 5) metrics. 
We also show that our adapted sequential channel and spatial attention structure is superior to its conventional parallel implementation. 

\section{conclusion}

In this work, we propose a novel feature-rich network (HDRfeat) for HDR reconstruction including a channel-wise feature extraction network to extract and bottleneck rich contextual information from multi-exposure images to fit in the HDR domain, a residual attention block with sequential channel and spatial attention to concentrate on features with large motions and an HDR reconstruction network with dilated residual feature distillation block (RFDB) as the backbone. We perform qualitative and quantitative comparisons on the public benchmark dataset, showing that the proposed method outperforms the state-of-the-art methods.  


\footnotesize{
\bibliographystyle{IEEEtran}
\balance
\bibliography{ref.bib}
}
\end{document}